\documentclass[12pt]{article}
\usepackage{epsfig}

\newcommand\GeV{\ifmmode {\mathrm{\ Ge\kern -0.1em V}}\else
                   \textrm{Ge\kern -0.1em V}\fi}%

\newcommand\pubnumber{ }
\newcommand\pubdate{March 3, 2001}
\newcommand\hepnumber{hep-ph/0103040}

\def\csumb{CERN, EP Division,  CH1211, Gen\`eve 23, 
Switzerland\footnote{On leave of absence from INFN Sezione di
    Napoli,  80125, Napoli, Italy.}}
\def\support{\footnote{e-mail: Salvatore.Mele@cern.ch.}} 

\def\Title#1{\begin{center} {\Large\bf #1 } \end{center}}
\def\Author#1{\begin{center}{ \sc #1} \end{center}}
\def\Address#1{\begin{center}{ \it #1} \end{center}}

\newcommand\pubblock{\rightline{\begin{tabular}{l} \pubnumber\\
         \pubdate\\ \hepnumber \end{tabular}}}
\newenvironment{Abstract}{\begin{quotation}  }{\end{quotation}}
\newenvironment{Presented}{\begin{quotation} \begin{center} 
             Presented at the\end{center}
      \begin{center}\begin{large}}{\end{large}\end{center} \end{quotation}}
\def\Acknowledgments{\bigskip  \bigskip \begin{center}
          \large\bf Acknowledgments\end{center}}

\makeatletter
\def\section{\@startsection{section}{0}{\z@}{5.5ex plus .5ex minus
 1.5ex}{2.3ex plus .2ex}{\large\bf}}
\def\subsection{\@startsection{subsection}{1}{\z@}{3.5ex plus .5ex minus
 1.5ex}{1.3ex plus .2ex}{\normalsize\bf}}
\def\subsubsection{\@startsection{subsubsection}{2}{\z@}{-3.5ex plus
-1ex minus  -.2ex}{2.3ex plus .2ex}{\normalsize\sl}}

\renewcommand{\@makecaption}[2]{%
   \vskip 10pt
   \setbox\@tempboxa\hbox{\small #1: #2}
   \ifdim \wd\@tempboxa >\hsize     
       \small #1: #2\par          
     \else                        
       \hbox to\hsize{\hfil\box\@tempboxa\hfil}
   \fi}

 \def\citenum#1{{\def\@cite##1##2{##1}\cite{#1}}}
 
\newcount\@tempcntc
\def\@citex[#1]#2{\if@filesw\immediate\write\@auxout{\string\citation{#2}}\fi
  \@tempcnta\z@\@tempcntb\m@ne\def\@citea{}\@cite{\@for\@citeb:=#2\do
    {\@ifundefined
       {b@\@citeb}{\@citeo\@tempcntb\m@ne\@citea\def\@citea{,}{\bf ?}\@warning
       {Citation `\@citeb' on page \thepage \space undefined}}%
    {\setbox\z@\hbox{\global\@tempcntc0\csname b@\@citeb\endcsname\relax}%
     \ifnum\@tempcntc=\z@ \@citeo\@tempcntb\m@ne
       \@citea\def\@citea{,}\hbox{\csname b@\@citeb\endcsname}%
     \else
      \advance\@tempcntb\@ne
      \ifnum\@tempcntb=\@tempcntc
      \else\advance\@tempcntb\m@ne\@citeo
      \@tempcnta\@tempcntc\@tempcntb\@tempcntc\fi\fi}}\@citeo}{#1}}
\def\@citeo{\ifnum\@tempcnta>\@tempcntb\else\@citea\def\@citea{,}%
  \ifnum\@tempcnta=\@tempcntb\the\@tempcnta\else
  {\advance\@tempcnta\@ne\ifnum\@tempcnta=\@tempcntb \else\def\@citea{--}\fi
    \advance\@tempcnta\m@ne\the\@tempcnta\@citea\the\@tempcntb}\fi\fi}
\makeatother

\def\beq{\begin{equation}}
\def\eeq#1{\label{#1}\end{equation}}
\def\eeqn{\end{equation}}

\newenvironment{Eqnarray}%
   {\arraycolsep 0.14em\begin{eqnarray}}{\end{eqnarray}}
\def\beqa{\begin{Eqnarray}}
\def\eeqa#1{\label{#1}\end{Eqnarray}}
\def\eeqan{\end{Eqnarray}}

\let\bar=\overbar

\def\Dslash{\not{\hbox{\kern-4pt $D$}}}
\def\dslash{\not{\hbox{\kern-2pt $\del$}}}

\def\msb{{\bar{\ssstyle M \kern -1pt S}}}

\def\lsim{\mathrel{\raise.3ex\hbox{$<$\kern-.75em\lower1ex\hbox{$\sim$}}}}
\def\gsim{\mathrel{\raise.3ex\hbox{$>$\kern-.75em\lower1ex\hbox{$\sim$}}}}

\begin{document}
\begin{titlepage}
\pubblock

\vfill
\def\thefootnote{\fnsymbol{footnote}}
\Title{Indirect Determination of the Vertex and Angles\\[5pt] 
 of the Unitarity Triangle}
\vfill
\Author{Salvatore Mele\support}
\Address{\csumb}
\vfill
\begin{Abstract}
The values of the elements of the Cabibbo-Kobayashi-Maskawa 
matrix are constrained by direct and indirect measurements.
A fit to  experimental data and  theory calculations allows 
the indirect determination of the  vertex and angles of the
unitarity triangle  as:
\[
    \rho =0.18 \pm 0.07\,\,\,\,\,
    \eta =0.35 \pm 0.05
\]
\[
    \sin{2\alpha}  =0.14 ^{+0.25}  _{-0.38} \,\,\,\,\,
    \sin{2\beta}   =0.73 \pm 0.07 \,\,\,\,\,
    \gamma         =63 ^{+\phantom{0}8}  _{-11}\,\rm degrees.
\]
Information is derived on the presence of CP violation in the matrix,  on
non-per\-tur\-ba\-ti\-ve QCD parameters and  
on the $\rm B^0_s$ oscillation frequency.
\end{Abstract}
\vfill
\begin{Presented}
5th International Symposium on Radiative Corrections \\ 
(RADCOR--2000) \\[4pt]
Carmel CA, USA, 11--15 September, 2000
\end{Presented}
\vfill
\end{titlepage}
\def\thefootnote{\arabic{footnote}}
\setcounter{footnote}{0}
%
%
%

\section{Introduction}

In the Standard Model of the electroweak
interactions~\cite{Glashow:1961tr,Salam:ed.rm,Weinberg:1967tq}, a $3\times3$ unitary matrix
describes the
mixing of the quark mass eigenstates into the weak interaction ones.
This  matrix is known as the
Cabibbo-Kobayashi-Maskawa  (CKM) matrix~\cite{Cabibbo:1963yz,Kobayashi:1973fv}, and can be written in
terms of just four real parameters~\cite{Wolfenstein:1983yz}:
\begin{equation}
  \mathrm{
    V_{CKM} = \pmatrix{ \mathrm{V_{ud}} &  \mathrm{V_{us}} &  \mathrm{V_{ub}} \cr 
      \mathrm{V_{cd}} &  \mathrm{V_{cs}} &  \mathrm{V_{cb}} \cr 
      \mathrm{V_{td}} &  \mathrm{V_{ts}} &  \mathrm{V_{tb}} \cr}
    }
  \simeq \pmatrix
  {
    1-{\lambda^2 \over 2} & \lambda &  A\lambda^3 (\rho - i\eta) \cr
    -\lambda & 1-{\lambda^2 \over 2} &  A\lambda^2 \cr
    A\lambda^3 (1-\rho - i\eta) &  -A\lambda^2 &      1 \cr
  }.
\label{equation:ckm}
\end{equation}   
$A$, $\rho$ and $\eta$ are of  order  unity and
$\lambda$ is  the sine of the Cabibbo angle. The  parameter
$\eta$ is the complex phase of the matrix, directly related to the
violation of the CP symmetry 
in the weak interactions.
The measurement of the parameters of the CKM matrix is of 
fundamental importance for both the  description of the
weak interaction of quarks and to shed light on
the mechanism of CP violation. 

\begin{figure}[hb!]
\begin{center}
\epsfig{file=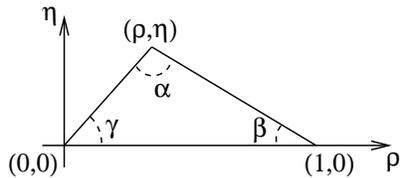,height=0.9in}
\caption[0]{\label{fig:1}The unitarity triangle.}
\end{center}
\end{figure}

The parameters $A$ and $\lambda$ are known with an accuracy of a few
percent and this work concentrates on the indirect determination
of $\rho$ and $\eta$. 
This 
is also described as the study of the vertex or the angles
of a triangle in the $\rho-\eta$ plane, whose other two vertices are
located in (0,0) and (1,0). 
This triangle, called the unitarity triangle, is depicted in
Figure~\ref{fig:1}. This study follows the same procedure as a
previous publication~\cite{Mele:1999bf} with an update of the input
parameters, as described in the following.

A large number
of physical processes are parametrised in terms of the values of
the elements of the CKM matrix.
Among them, four present the largest sensitivity to $\rho$ and $\eta$, given the
knowledge of the involved theoretical and
experimental quantities. These processes
 are discussed in the following and then used in a fit
to derive  $\rho$ and $\eta$. Conclusions are then drawn from the
results of this fit.

%
%

\section{Constraints}

\subsection{$\lambda$ and $A$}

The value of the sine of the Cabibbo angle  is measured
as~\cite{Groom:2000in}:  
\[
\lambda = 0.2196 \pm 0.0023.
\]
The study of inclusive semileptonic B decays by the 
CLEO~\cite{Barish:1996cx}  
and the LEP~\cite{Abbaneo:2000ej} experiments yields information on the value of
$|\mathrm{V_{cb}}|$. Further constraints are derived from the study of
the $\rm B^0 \rightarrow D^{*+}\ell \nu$ decay, both at the
$\Upsilon(4\rm S)$~\cite{Cinabro:2000rj} and at the Z
pole~\cite{Abbaneo:2000ej}.
From these measurements, a value
$|\mathrm{V_{cb}}| = (40.9 \pm 1.9)\times 10^{-3}$
is extracted, which yields:
\[
A = {|\mathrm{V_{cb}}| \over \lambda^2} = 0.83 \pm 0.04.
\]

%
%

\subsection{CP Violation for Neutral Kaons}

The mass eigenstates of the
neutral kaons can be written as
$|{\rm K_S}\rangle = p |{\rm K^0}\rangle + q |\bar{\rm K}^0\rangle$ and 
$|{\rm K_L}\rangle = p |{\rm K^0}\rangle - q |\bar{\rm K}^0\rangle$.
The relation $p\neq q$ implies the violation of the CP symmetry that, in the 
Wu-Yang phase convention~\cite{Wu:1964qx}, is described by the parameter $\epsilon_K$
defined as:
\[
{p \over q} = {1 + \epsilon_K \over 1 - \epsilon_K  }.
\]
The precise measurements of the $\rm K_S \rightarrow \pi^+ \pi^-$ and 
$\rm K_L \rightarrow \pi^+ \pi^-$ decay rates imply~\cite{Groom:2000in}:
\[
|\epsilon_K| = (2.280 \pm 0.019) \times 10^{-3}.
\]
The relation of $|\epsilon_K|$ to the CKM matrix parameters 
is~\cite{Buras:1984ap,Buras:1984pq}:
\begin{eqnarray}
|\epsilon_K| & = & {G_F^2 f^2_K m_K m_W^2 \over 6 \sqrt{2} \pi^2 \Delta m_K}
 B_K \left( A^2 \lambda^6 \eta \right) \\ \nonumber
& \times & \big[ y_c \left( \eta_{ct}f_3(y_c,y_t)-  \eta_{cc}\right) 
 + \eta_{tt} y_t f_2(y_t) A^2 \lambda^4 \left(1-\rho\right)\big].                          
\end{eqnarray}
The functions $f_3$ and $f_2$ of the variables
$y_t = m_t^2 / m_W^2$ and $y_c = m_c^2 / m_W^2$ are given
in Reference~\cite{Ali:1995es}.
The measured value of the top quark mass, $174.3 \pm
5.1\,\GeV$~\cite{Groom:2000in}, is scaled as proposed
in Reference~\cite{Buras:1990fn}, giving: 
\[
\overline{m_t}(m_t) = 167.3 \pm 5.2\,\GeV,
\]
while the mass of the charm quark is chosen as~\cite{Groom:2000in}:
\[
\overline{m_c}(m_c) = 1.25 \pm 0.10\,\GeV.
\]
The calculated QCD corrections in Equation~(2) are described by the
consistent set of parameters~\cite{Buras:1990fn,Herrlich:1994yv,Herrlich:1995hh,Herrlich:1996vf}:
\[
\eta_{cc} = 1.38 \pm 0.53,\,\,\,\,
\eta_{tt} = 0.574 \pm 0.004,\,\,\,\,
\eta_{ct} = 0.47 \pm 0.04.
\]
Non-per\-tur\-ba\-ti\-ve QCD contributions to this process are affected by a
large uncertainty and are summarised by the ``bag'' 
parameter $B_K$, chosen as~\cite{Kuramashi:2000gt}:
\[
B_K  = 0.87\pm 0.14.
\]
The other physical constants appearing in Equation~(2)  are reported in
Table~1. 
The measurement of $|\epsilon_K|$ constrains the vertex of the
unitarity triangle onto an hyperbola in the 
the $\rho-\eta$ plane.

Recent measurements of 
direct CP violation in the neutral kaon sector from the
KTEV~\cite{Alavi-Harati:1999xp} and NA48~\cite{Fanti:1999nm} experiments 
confirm the
previous NA31 result~\cite{Barr:1993rx}. These measurements could result in a
lower bound to $\eta$. Nonetheless they are  not used to 
constrain the CKM matrix owing to the large uncertainties that
affect
the corresponding theoretical
calculations~\cite{Bertolini:2000uj,Buras:1998ra}. 

%
%

\subsection{Oscillations of $\boldmath{\rm B^0_d}$ Mesons}

The behaviour of neutral mesons containing a $b$ quark 
depends on the 
the mass difference 
between the heavy and light mass eigenstates, $\rm B_H$ and $\rm B_L$. These are
different from the CP eigenstates $\rm B^0_d$ and $\rm \bar{B}^0_d$.
The mass difference, $\Delta m_d = m_{\rm{B_H}} -  m_{\rm{B_L}}$, is
measured~\cite{Stocchi:2000ps} at LEP, the $\Upsilon(\rm 4S)$ and the
TEVATRON by means of the study of the oscillations 
of one CP eigenstate into the other. A recent average is:
\[
\Delta m_d            = 0.487 \pm 0.014\,\mathrm{ps}^{-1}.
\]
Recent results from the Babar~\cite{Touramanis:2001wc} and Belle~\cite{Abe:2000yh}
collaborations are not yet included in this average with which they are statistically comparable.
The value of $\Delta m_d$ is related to the CKM parameters as:
\begin{equation}
\Delta m_d = {G_F^2 \over 6 \pi ^2}m_W^2 m_B \left(f_{B_d}\sqrt{B_{B_d}}\right)^2
\eta_B y_t f_2(y_t) A^2 \lambda^6 \left[ \left(1-\rho\right)^2 + \eta^2\right].
\label{equation:dmd}
\end{equation}
The
calculated QCD correction $\eta_B$ amounts to~\cite{Buras:1990fn,Herrlich:1994yv,Herrlich:1995hh,Herrlich:1996vf}:
\[
     \eta_B  = 0.55 \pm 0.01,
\]
while non-per\-tur\-ba\-ti\-ve QCD contributions are summarised by~\cite{Becirevic:2000nv}:
\[
     f_{B_d}\sqrt{B_{B_d}} = 0.206 \pm 0.029\,\GeV.
\]

The vertex of the unitarity triangle is constrained by  $\Delta m_d$
onto a circle in the $\rho-\eta$ plane, with centre in $(1,0)$.


\subsection{Oscillations of $\boldmath{\rm B^0_s}$ Mesons}

The $\rm B^0_s$ mesons are predicted to mix like
the  $\rm B^0_d$ mesons, but their larger mass difference,
$\Delta m_s$, results into faster
oscillations. These have eluded direct observation and the
 current 95\% Confidence Level (CL) lower limit on $\Delta m_s$ from the  
LEP,  SLD and CDF collaborations is~\cite{Stocchi:2000ps}:
\[
  \Delta m_s > 14.9\,\mathrm{ps}^{-1}\,\,\, (95\%\,\mathrm{CL}).
\]
The experiments, once combined, are sensitive to values of $\Delta
m_s$ up to $17.9\,\mathrm{ps}^{-1}$ and a 2.5\,$\sigma$ indication for
the observation of $\rm B^0_s$  oscillations is observed
around $ \Delta m_s = 17.7 \,\mathrm{ps}^{-1}$.

The expression for $\Delta m_s$ as a function of the CKM parameters
is similar to that 
for  $\Delta m_d$, and taking their  ratio, it follows:
\begin{equation}
\Delta m_s = \Delta m_d {1 \over \lambda^2}{m_{B_s} \over m_{B_d}} \xi^2 
{1 \over \left( 1- \rho\right)^2 + \eta^2}.
\label{equation:dms}
\end{equation}
All the theoretical parameters and their uncertainties are included in the  quantity
$\xi$, known as~\cite{Becirevic:2000nv}:
\[
\xi = { f_{B_d}\sqrt{B_{B_d}} \over f_{B_s}\sqrt{B_{B_s}} } = 1.16 \pm 0.07.
\]

The  lower limit on  $\Delta m_s$ constrains the vertex
of the unitarity triangle in a circle in the $\rho-\eta$ plane  with centre in $(1,0)$.


\subsection{Charmless Semileptonic b Decays}

The constraints described so far suffer from the uncertainties in 
non-pertur\-ba\-ti\-ve QCD quantities entering their expressions.
The determination of either 
$|\mathrm{V_{ub}}|$ or the ratio $|\mathrm{V_{ub}}|/|\mathrm{V_{cb}}|$ 
constitutes a constraint free from these uncertainties as:
\begin{equation}
 |\mathrm{V_{ub}}|/|\mathrm{V_{cb}}| = \lambda \sqrt{\rho^2 + \eta^2}.
\label{equation:vubvcb}
\end{equation}
The CLEO collaboration has measured this ratio by means
of the endpoint of inclusive~\cite{Bartelt:1993xh}  charmless semileptonic B decays as:
$|\mathrm{V_{ub}}|/|\mathrm{V_{cb}}| = 0.08 \pm 0.02$.
The ALEPH~\cite{Barate:1999vv}, DELPHI~\cite{Abreu:2000mx} and
L3~\cite{Acciarri:1998if}  collaborations  have
measured at LEP the  inclusive charmless semileptonic branching
fraction of beauty hadrons; these are averaged~\cite{Abbaneo:2000ej} as:
\[
|\mathrm{V_{ub}}| = (4.13  \,^{+0.63}_{-0.75})\times 10^{-3}.
\]
Using the quoted value of $|\mathrm{V_{cb}}|$, a combination with the CLEO measurement
yields:
\[
|\mathrm{V_{ub}}|/|\mathrm{V_{cb}}| = 0.089 \pm 0.010.
\]

\begin{figure}[htb!]
\begin{center}
\epsfig{file=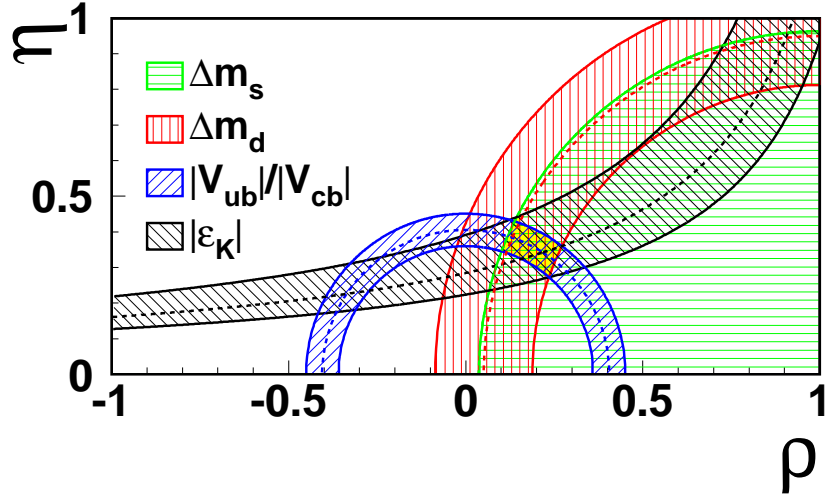,height=3in}
\caption[1]{\label{fig:2} Constraints in the $\rho-\eta$ plane.
      $\mathrm{B^0_s}$ oscillations are reported as a  95\% CL
      limit, while the other constraints represent a $\pm 1 \sigma$ variation of the experimental
      and theoretical parameters entering in the formulae in the
      text. Central values are indicated by the dashed lines. A darker
      area shows the overlap among the constraints.}
\end{center}
\end{figure}

This constraint is represented by a  circle in the $\rho-\eta$ plane
with centre in (0,0), 
shown in Figure~\ref{fig:2}, that also presents all the other constraints
described so far.

\begin{table}[ht!]
\caption{\label{tab:1}
Physical constants and parameters of the fit. The values not discussed
     in the text follow from
     Reference~\protect\cite{Groom:2000in}.}

\begin{center}
    \begin{tabular}{|rcl|rcl|} 
     \hline
     $\lambda               $& = & $0.2196(23)$ &                                 $A                     $& = & $0.83(4)$   \\               
     $G_F                   $& = & $1.16639(1)\times 10^{-5}$\,\GeV$^{-2}$&   $\eta_{ct}             $& = & $0.47(4)$   \\                
     $f_K                   $& = & $0.1598(15)$\,\GeV &                           $\eta_{cc}             $& = & $1.38(53)$  \\                 
     $\Delta m_K            $& = & $0.5304(14)\times 10^{-2}$\,ps$^{-1}$ &      $\overline{m_c}(m_c)   $& = & $1.25(10)$\,\GeV   \\          
     $m_K                   $& = & $0.497672(31)$\,\GeV &                       $\overline{m_t}(m_t)   $& = & $167.3(5.2)$\,\GeV  \\           
     $m_W                   $& = & $80.419(38)$\,\GeV &                            $f_{B_d}\sqrt{B_{B_d}} $&=& $0.206(29)$\,\GeV  \\           
     $m_{B_d}               $& = & $5.2792(18)$\,\GeV &                           $B_K                   $&=& $0.87(14)$ \\                    
     $m_{B_s}               $& = & $5.3692(20)$\,\GeV &                           $\xi                   $&=& $1.16(7)$ \\                    
     $m_B                   $& = & $5.290(2)$\,\GeV &                             $|\epsilon_K|          $& = & $2.280(19)\times 10^{-3}$  \\
     $\eta_B                $& = & $0.55(1)$ &                                     $\Delta m_d            $& = & $0.487(14)$\,ps$^{-1}$ \\     
     $\eta_{tt}             $& = & $0.574(4) $&                                   $|\mathrm{V_{ub}}|/|\mathrm{V_{cb}}| $& = & $0.089(10)$ \\ 
     \hline
   \end{tabular}
 \end{center}
\end{table}

%
%

\section{The fit}

The $\rho$ and $\eta$ parameters are determined from a 
fit to the constraints described above.
The experimental and theoretical quantities appearing in the  
formulae~(2),~(\ref{equation:dmd}), (\ref{equation:dms}) 
and~(\ref{equation:vubvcb})
are divided in two classes. Those whose uncertainties
are below 2\% are fixed to their central value as
listed in the left half of Table~1. The quantities affected by a larger
uncertainty and $|\epsilon_K|$ are considered as additional parameters of the fit, constraining
their values to the estimates summarised in the right half of Table~1.
The following expression is then minimised:
\begin{eqnarray*}
\chi^2 =  {\left(\widehat{A} - A\right)^2 \over \sigma_{A}^2} +   
          {\left(\widehat{m_c} - m_c\right)^2 \over \sigma_{m_c}^2} +
          {\left(\widehat{m_t} - m_t\right)^2 \over \sigma_{m_t}^2} + 
          {\left(\widehat{B_K} - B_K\right)^2 \over \sigma_{B_K}^2} +
          {\left(\widehat{\eta_{cc}} -\eta_{cc} \right)^2 \over \sigma_{\eta_{cc}}^2} + 
\\
          {\left(\widehat{\eta_{ct}} -\eta_{ct} \right)^2 \over \sigma_{\eta_{ct}}^2} + 
           {\left(\widehat{f_{B_d}\sqrt{B_{B_d}}} - f_{B_d}\sqrt{B_{B_d}}\right)^2 
                 \over \sigma_{f_{B_d}\sqrt{B_{B_d}}}^2} +
           {\left(\widehat{\xi} - \xi\right)^2 \over \sigma_{\xi}^2} + 
           {\left(\widehat{|\mathrm{V_{ub}}|\over|\mathrm{V_{cb}}|} 
               - {|\mathrm{V_{ub}}|\over|\mathrm{V_{cb}}|}\right)^2 \over \
             \sigma_{{|\mathrm{V_{ub}}| \over |\mathrm{V_{cb}}|}}^2} +
\\
          {\left(\widehat{|\epsilon_K|} - |\epsilon_K| \right)^2  
          \over  \sigma_{|\epsilon_K|}^2} +
           {\left(\widehat{\Delta m_d} - \Delta m_d\right)^2
          \over  \sigma_{\Delta m_d}^2}+
          \chi^2\left( {\cal{A}}\left(\Delta m_s\right), \sigma_{\cal{A}}\left(\Delta m_s \right)\right).
\end{eqnarray*}
The symbols with a hat represent the reference values and the corresponding $\sigma$
denote their uncertainties. 
The free parameters of the fit are $\rho$, $\eta$, $A$, $m_c$, $m_t$, $B_K$, $\eta_{ct}$, 
$\eta_{cc}$, $f_{B_d}\sqrt{B_{B_d}}$ and $\xi$, used to calculate
the values of $|\epsilon_K|$,  $\Delta m_d$,  $\Delta m_s$ and
$|\mathrm{V_{ub}}| \ |\mathrm{V_{cb}}|$
by means of the formulae~(2),~(\ref{equation:dmd}),~(\ref{equation:dms}) 
and~(\ref{equation:vubvcb}).

As $\Delta m_s$ is not yet measured, its experimental information
has to be included in the
$\chi^2$ 
following a different approach~\cite{Mele:1999bf}.
The results of the search for $\rm B^0_s$ oscillations are
combined~\cite{Stocchi:2000ps} 
in terms of the oscillation amplitude ${\cal{A}}$~\cite{Moser:1997xf},
a parameter that is zero in the absence of any signal and compatible with
one otherwise, as expressed by the oscillation probability $P$:
\[
P \left[ \rm B^0_s \rightarrow (B^0_s, \bar{B}^0_s) \right]
= {1 \over 2 \tau_s} e^{-t/\tau_s} \left( 1 \pm {\cal {A}} \cos{\Delta m_s}
\right).
\]
The results of different experiments are combined  in terms of  ${\cal{A}}\left(\Delta m_s\right)$
and of its uncertainty $\sigma_{\cal{A}}\left(\Delta m_s \right)$. The
95\% CL limit on  $\Delta m_s$  is the value for which the area above
one of the Gaussian distribution 
with mean ${\cal{A}}\left(\Delta m_s\right)$ and variance
$\sigma^2_{\cal{A}}\left(\Delta m_s \right)$ 
equals 5\% of its total area.
In the fit, each value taken by the parameters  $\rho$, $\eta$ and $\xi$ is
converted into a value of $\Delta m_s$ by means of
formula~(\ref{equation:dms}). A value of the CL for the
oscillation hypothesis is then calculated by integrating 
the Gaussian distribution
with mean ${\cal{A}}\left(\Delta m_s\right)$ and variance
$\sigma^2_{\cal{A}}\left(\Delta m_s \right)$. The value  
$\chi^2\left( {\cal{A}}\left(\Delta m_s\right), \sigma_{\cal{A}}\left(\Delta m_s \right)\right)$
of a $\chi^2$ distribution with one degree of freedom corresponding to
this CL is then 
calculated and finally added to the $\chi^2$ of the fit.

\begin{figure}[htb!]
\begin{center}
\epsfig{file=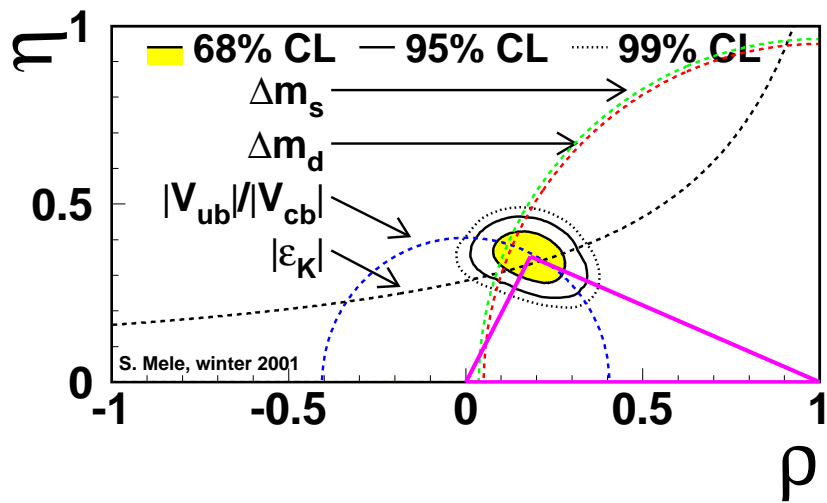,height=3in}
\caption[1]{\label{fig:3} The favoured unitarity triangle
      and the confidence regions for its vertex;  the $\Delta m_s$ limit and the central 
      values of the other constraints are also shown.}
\end{center}
\end{figure}

The fit indicates  the following values for the $\rho$ and $\eta$ parameters:
\[
    \rho =0.18 \pm 0.07\,\,\,\,\,
    \eta =0.35 \pm 0.05
\]
\[
     0.05 < \rho < 0.30 \,\,\,\,\,
     0.26 < \eta < 0.44 \,\,\,(\mathrm{95\% CL}).
\]
No large change in these results is observed if the theory contribution constraints are removed from the $\chi^2$ and
a flat
distribution within the uncertainties is used in their place.
Figure~\ref{fig:3} presents the confidence regions for the vertex of the
unitarity triangle.
The value of the angles
of the unitarity triangle are determined as as: 
\[
    \sin{2\alpha}  =0.14 ^{+0.25}  _{-0.38} \,\,\,\,\,
    \sin{2\beta}   =0.73 \pm 0.07 \,\,\,\,\,
    \gamma         =63 ^{+\phantom{0}8}  _{-11}\,\rm degrees.
\]
\[
   -0.77 <  \sin{2\alpha} < 0.50 \,\,\,\,\,
   0.59 <  \sin{2\beta}  < 0.87 \,\,\,\,\,
   44^\circ  <   \gamma       < 82^\circ  \,\,\,(\mathrm{95\% CL})  
\]
The angles $\alpha$ and $\beta$ are reported in terms of the
 functions $\sin{2\alpha}$ and $\sin{2\beta}$, to which
 the studies of the CP symmetry usually refer.

Several direct measurements of $\sin{2\beta}$ were recently reported,
as listed in Table~2. Their average, $ \sin{2\beta}^{\rm exp.}   =0.48
\pm 0.16 $, is lower but in agreement with the present estimate,
that does not make use of this direct information.

\begin{table}[ht!]
\caption{\label{tab:2} Measurements of $ \sin{2\beta}$ and their
  average, compared with the fit result.} 

\begin{center}
    \begin{tabular}{|rl|} 
     \hline
\rule{0pt}{12pt}  Aleph~\protect\cite{Barate:2000tf}  &  $ 0.93 ^{+0.64\,\,+0.36}  _{-0.88\,\,-0.24}$ \\
\rule{0pt}{12pt}  BaBar~\protect\cite{Aubert:2001sp}          &  $ 0.34 \pm 0.20 \pm 0.05$ \\
\rule{0pt}{12pt}  Belle~\protect\cite{Abashian:2001pa}&  $ 0.58 ^{+0.32\,\,+0.09}  _{-0.34\,\,-0.10}$ \\
\rule{0pt}{12pt}  CDF~\protect\cite{Affolder:2000gg}  &  $0.79 ^{+0.41}  _{-0.44}$ \\
  \hline
\rule{0pt}{12pt} Average & $ 0.48 \pm 0.16$ \\
\rule{0pt}{12pt} This fit     & $0.73 \pm 0.07$ \\
     \hline
   \end{tabular}
 \end{center}
\end{table}

%
%

\section{Consequences of the fit}

\begin{figure}[hb!]
\begin{center}
\epsfig{file=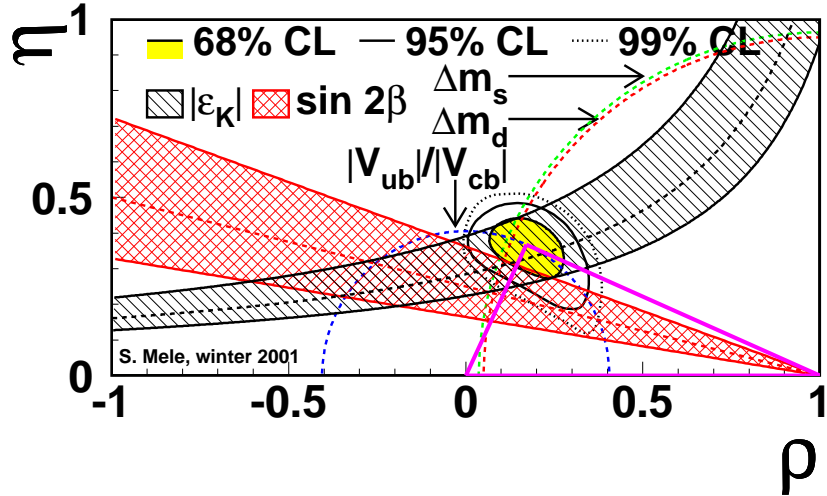,height=3in}
\caption[1]{\label{fig:4} The favoured unitarity triangle
      and the confidence regions for its vertex, no information from
      the kaon system is used in the fit. The
      experimental measurements of CP violation in the 
      neutral kaon and neutral $b$ meson systems are superimposed.}
\end{center}
\end{figure}

A strong experimental evidence for CP violation in
the CKM matrix, described by values of its complex phase,
$\eta$, different from zero, comes from the neutral kaon system.
It is of interest~\cite{Barbieri:1998bm} to investigate whether processes other than kaon
physics predict a
value of $\eta$
compatible with zero or not. 
Figure~\ref{fig:4} presents the results of a fit from which the information from
the kaon system is removed.. The presence of a CP
violating phase in the matrix, {\it i.e.} its 
complex nature is strongly favoured. 

The values of the parameters $B_K$ and $f_{B_d}\sqrt{B_{B_d}}$  that describe non-per\-tur\-ba\-ti\-ve QCD effects
can be estimated by removing their constraint  from the fit. This
procedure yields:
\[
    \rho =0.17 \pm 0.07 \,\,\,\,\,
    \eta = 0.38 ^{+0.05} _{-0.06} \,\,\,\,\, 
    B_K  = 0.76 ^{+0.21} _{-0.15}
\]
in the first case and in the second:
\[
    \rho =0.21 ^{+0.07} _{-0.08}  \,\,\,\,\,
    \eta = 0.34 \pm 0.05 \,\,\,\,\,
    f_{B_d}\sqrt{B_{B_d}} = 0.227 ^{+0.019} _{-0.015}\,\GeV.
\]
The fit indicates a value of $B_K$ with an uncertainty larger than the
input one, yet of similar magnitude.
The value of $f_{B_d}\sqrt{B_{B_d}}$ is found to be well in agreement
with the predicted one with a smaller uncertainty, this implies
that the high experimental precision of the $\Delta m_d$ constraint 
is not fully exploited by the
fit, limited by the uncertainty on the $f_{B_d}\sqrt{B_{B_d}}$ parameter.

The $\Delta m_s$ constraint heavily affects  the  $\rho$ 
uncertainty. Indeed, a fit that does not make use of the $\Delta m_s$
information results in:
\[
    \rho =0.09 _{-0.15} ^{+0.10}\,\,\,\,\,
    \eta =0.40 \pm 0.05.
\]
This fit is used to estimate the favoured values of  $\Delta m_s$ as:
\begin{eqnarray*}
    \Delta m_s  = 14.0 _{-3.3} ^{+3.4}\,\mathrm{ps}^{-1} &  \\
     7.4\,\mathrm{ps}^{-1} < \Delta m_s   < 21.0\,\mathrm{ps}^{-1} & \,\,\,(\mathrm{95\% CL}).
\end{eqnarray*}

%
%

\section{Conclusions}

The  measurements of $|\epsilon_K|$, $\Delta m_d$ and
$|\mathrm{V_{ub}}|$, together with the lower limit
on $\Delta m_s$, effectively constrain the CKM matrix:
from a fit to the experimental results and 
theory parameters, 
the vertex and the angles of the unitarity triangle are determined as:
\[
    \rho =0.18 \pm 0.07\,\,\,\,\,
    \eta =0.35 \pm 0.05
\]
\[
    \sin{2\alpha}  =0.14 ^{+0.25}  _{-0.38} \,\,\,\,\,
    \sin{2\beta}   =0.73 \pm 0.07 \,\,\,\,\,
    \gamma         =63 ^{+\phantom{0}8}  _{-11}\,\rm degrees.
\]
These results are in agreement with those of recent similar
analyses~\cite{Bargiotti:2000dn,Faccioli:2000mg,Ali:2000hy,Ciuchini:2000de,Buras:2001pn}.

A coherent picture of the current understanding of the CKM matrix is
presented by a fit that does not use any constraint from the kaon
system. Its results are displayed in Figure~\ref{fig:4}. The favoured region
for the vertex of the unitarity 
triangle corresponds to that experimentally indicated by the
measurement of the CP violation in the neutral kaon system and overlaps with
the one indicated by the recent measurements of $\sin{2\beta}$.


\Acknowledgments
I am grateful to the organisers of RADCOR-2000 for their
invitation to this interesting symposium.

\end{document}